\documentstyle[psfig]{mn}

\title[\Halpha\ star-formation rate at $z=1$] {Measurement of Star-Formation
Rate from \HalphaB\ in field galaxies at z=1 } 

\author[K. Glazebrook \etal]  {Karl Glazebrook,$^1$ Chris Blake,$^2$ 
Frossie Economou,$^3$ Simon Lilly,$^4$ Matthew Colless$^5$ \\
$^1$ Anglo-Australian Observatory, PO Box 296, Epping, NSW 2121, AUSTRALIA\\
$^2$ Magdalen College, Oxford, OX1 4AU, UK\hfill\\
$^3$ Joint Astronomy Centre, 660 North A'ohoku Place, University Park,
Hilo, HI 96720, USA \hfill\\
$^4$ Department of Astronomy, University of Toronto, 60 St George Street,
Toronto, Ontario, M5S 3H8, Canada \hfill\\
$^5$ Research School of Astronomy \& Astrophysics,
 The Australian National University, Weston Creek, ACT 2611, Australia \hfill\\
}

\pagerange{\pageref{firstpage}--\pageref{lastpage}}
\pubyear{1998}

\def\Msun{\hbox{$M_{\odot}$}}
\def\Zsun{\hbox{$Z_{\odot}$}}
\def\Hunits{\hbox{$\rm km\,s^{-1}\,Mpc^{-1}$}}

\def\etal{{\it et~al.\/}}
\def\gs{\mathrel{\lower0.6ex\hbox{$\buildrel {\textstyle >}
 \over {\scriptstyle \sim}$}}}
\def\ls{\mathrel{\lower0.6ex\hbox{$\buildrel {\textstyle <}
 \over {\scriptstyle \sim}$}}}
\def\micron{\hbox{$\mu\rm m$}}

\def\Halpha{\hbox{$\rm H\alpha$}}
\def\HalphaB{\hbox{$\bf\boldmath H\alpha$}}
\def\Hbeta{\hbox{$\rm H\beta$}}
\def\HUV{\hbox{UV$/\Halpha$}}

\begin{document}

\label{firstpage}

\maketitle

\begin{abstract}

We report the results of $J$-band infrared spectroscopy of a sample of
13 $z=1$ field galaxies drawn from the Canada-France Redshift Survey,
targeting galaxies whose redshifts place the rest frame \Halpha\ line
emission from HII regions in between the bright night sky OH lines. As
a result we detect emission down to a flux limit of $\simeq 10^{-16}$
ergs cm$^{-2}$ s$^{-1}$ corresponding to a luminosity limit of $\simeq
10^{41}$ ergs at this redshift for a $H_0=50$ km s$^{-1}$ Mpc,$^{-1}$
$q_0=0.5$ cosmology. From these luminosities we derive estimates of
the star-formation rates in these galaxies which are independent of
previous estimates based upon their rest-frame ultraviolet (2800\AA)
luminosity. The mean star-formation rate at $z=1$, from this
sample, is found to be at least three
times as high as the ultraviolet estimates. The dust extinction in
these galaxies is inferred to be moderate, for
standard extinction laws, with a typical $A_V=0.5$--1.0
mags, comparable to local field galaxies. This suggests that the bulk of 
star-formation is not heavily obscured, unless one uses greyer extinction
laws.

Star-forming galaxies have the bluest colours and a
preponderance of disturbed/interacting morphologies. We also
investigate the effects of particular star-formation histories, in
particular the role of bursts vs continuous star-formation in changing
the detailed distribution of UV to \Halpha\ emission. Generally we
find that models dominated by short, overlapping, bursts at typically
0.2 Gyr intervals provide a better model for the data than a
constant rate of star-formation.
The star-formation history of the Universe from Balmer lines is compiled
and found to be typically 2--3$\times$ higher than that inferred from the UV
{\em at all redshifts}. It can not yet be clearly established whether
the star-formation rate falls off or remains constant at high-redshift.

\end{abstract}


\section{INTRODUCTION}

The topic of the history of star-formation in the Universe has excited
much interest in recent years, stimulated by the first observations of
nearly-normal star-forming galaxies at $z>3$ (Steidel \etal\ 1996).
Previously high-redshift studies were limited to highly active galaxies that
may be poor tracers of the typical star-formation history of the
universe as a whole.
Steidel \etal\ used the colour signature of the Lyman
break/Lyman-$\alpha$ forest discontinuity being redshifted through
optical filters to select high-redshift objects (Guhathakurta \etal\ 1990),
these were subsequently confirmed spectroscopically on the
10m W.M. Keck telescope. 

Comparison of these objects with the low-redshift ($z<1$) samples of
field galaxies (Lilly \etal\ 1995, Ellis \etal, 1996) 
appears to show a rise in the
Universal star-formation rate from $z=0$ to $z=1$ and a drop-off at
$z>3$ indicating a star-formation peak in the $z=1$--2 epoch (Madau
\etal\ 1996). This is also inferred to be the epoch when large
galaxies with classical elliptical and spiral morphologies are
assembled: Hubble Space Telescope observations indicate they are
extant at $z=1$ (Brinchmann \etal\ 1998, Lilly \etal\ 1998) but absent
in the $z>3$ sample (Giavalisco \etal\ 1996, Lowenthal \etal\
1997). Theoretical developments using galaxy formation simulations
constrained by the observed evolution in the density of neutral gas
from Lyman-$\alpha$ QSO absorbers show qualitative agreement with this
picture (Fall \etal\ 1996).

However these measurements of star-formation rate are based upon the
measurement of ultraviolet continuum luminosity, 1500--2800\AA\ in the 
rest-frame, assumed to be from young stellar populations. If dust
extinction played a significant role in obscuring UV
radiation they could be under-estimated by
large factors which may change the picture completely.

A more robust way to measure the star-formation rate of high-redshift
galaxies would be to measure their luminosities in Balmer
recombination lines. This radiation come from reprocessed ionising
radiation emitted by young  
stars. This approach has two advantages: firstly the
ionising radiation comes from more massive short lived stars than the
softer 1500--2800\AA\ UV and hence falls quickly to zero only 20 Myr
after star-formation stops. Thus the Balmer luminosity is a more
direct measure of the {\em instantaneous} star-formation rate.  This
contrasts with the UV which continues to rise as the stellar
populations evolve, typically doubling for example between 10 and 1000
Myr at 1500\AA. For \Halpha\ the main dependence is
directly on the Initial Mass Function and negligibly on the temporal
evolution.

The second main advantage is that the Balmer radiation is emitted in
the red part of the optical spectrum and is thus much less affected by
any dust extinction or attenuation than the ultraviolet. For example
for typical SMC and Milky Way extinction laws (Pei 1992) the 1500\AA\
and 2800\AA\ extinctions (in magnitudes) range from 2--7 times greater
than that at \Halpha\ (6563\AA).

However to observe \Halpha\ at high-redshift requires infrared
spectroscopy, which has not been possible until recently because of
the faintness of the sources involved. Pettini \etal\ (1998) have
secured the first IR spectra of 5 of the $z>3$ Steidel \etal\
galaxies, and obtained \Hbeta\ luminosities. In this paper we report
the results of the first measurements of the \Halpha\ line in a sample
of normal $z\simeq 1$ field galaxies drawn from the Canada-France
Redshift Survey (Lilly \etal\ 1995, Le Fevre \etal\ 1995, 
Lilly \etal\ 1995B, Hammer \etal\ 1995). This sample (hereafter `CFRS')
is a highly-complete redshift survey of a magnitude-selected 
($I_{AB}<22.5$) sample of normal field galaxies. The median redshift is
0.6, and galaxies extend out to $z=1.3$. Because the sample is magnitude
selected the $z\gs 1$ end is dominated by luminous $L\sim L^*$ galaxies.

\section{OBSERVATIONS AND DATA REDUCTION}

The observations were carried out on May 10-11 and October 3-5 1996 at
the UK Infrared Telescope in Hawaii using the CGS4 spectrograph
(Wright 1994). We chose galaxies in the redshift range 0.790-1.048 so
that the \Halpha\ line would lie in the relatively clean part of the
near-infrared $J$-band (1.17-1.34\,\micron) where the atmospheric
extinction is relatively low ($<15\%$). We also selected galaxies with
detectable [OII] emission in the optical CFRS spectra which make up
85\% of the CFRS sample at $z\sim 1$.

As well as absorption, the $J$-band is contaminated by numerous airglow
OH emission lines, which increase the broad-band sky-brightness by a
factor of 30 and hinders the detection of faint objects.  Our
observational strategy was to observe at high-resolution with CGS4,
thus resolving out the OH background. Because we already knew the
galaxies' redshifts from the optical spectra the limited wavelength
coverage at high-resolution was not a problem.  Moreover we could
exclude galaxies whose redshifts would put the \Halpha\ line on or
close to an OH line. (A similar strategy was also adopted by Pettini
\etal) We observed with the 150 lines/mm grating and the 1 arcsec/1
pixel slit giving a resolution $\lambda/\Delta\lambda=2200$. We
determined empirically that at this resolution OH lines contaminated
50\% of the bandpass, {\it i.e.}  we had to exclude 50\% of the
high-redshift galaxies, but in the remaining clean part of the
bandpass the mean background was only 20\% that of broad $J$.

Targets were acquired using the following procedure: first we
`peaked-up' on a very bright star ($V=1-2$ mags) within 1--2 degrees
of the target.  This involves centering the star on the optical finder
TV, reading the IR array in a continuous `MOVIE' mode, and then
adjusting the offset between the axis of CGS4 and the TV until the IR
flux is maximised. This assures the IR slit is aligned with the TV
crosshair.  Then we went to a fainter star, typically 17th mag, within
1--2 arcmin of our target galaxy and measured off the same coordinate
system, centered the star on the TV crosshair and did a blind offset
on to the target galaxy. (The targets were too faint to see on the
TV). We would then autoguide either on the offset star or another
bright star in the region.

Observations were made stepping the InSb detector array in 0.5 pixel
increments to fully sample the instrument profile and nodding the
telescope $\pm$ 9 arcsec along the slit between `OBJECT' and `SKY'
positions to facilitate sky-subtraction (though note the object is
still on the slit in both positions). Individual exposures ranged from
10-15 minutes. The typical seeing was 1.0 arcsec.  A total of 13
objects were observed: these are listed with their total exposure
times in Table~\ref{tab-observations}. Standard wavelength calibration
and flatfield corrections were applied. The October observing run was
affected by the spectrograph slit being jammed out of position which
caused the lines to be tilted on the image. The shifts were measured
by cross-correlation and the tilt corrected by re-interpolation.

\begin{figure*} 
\psfig{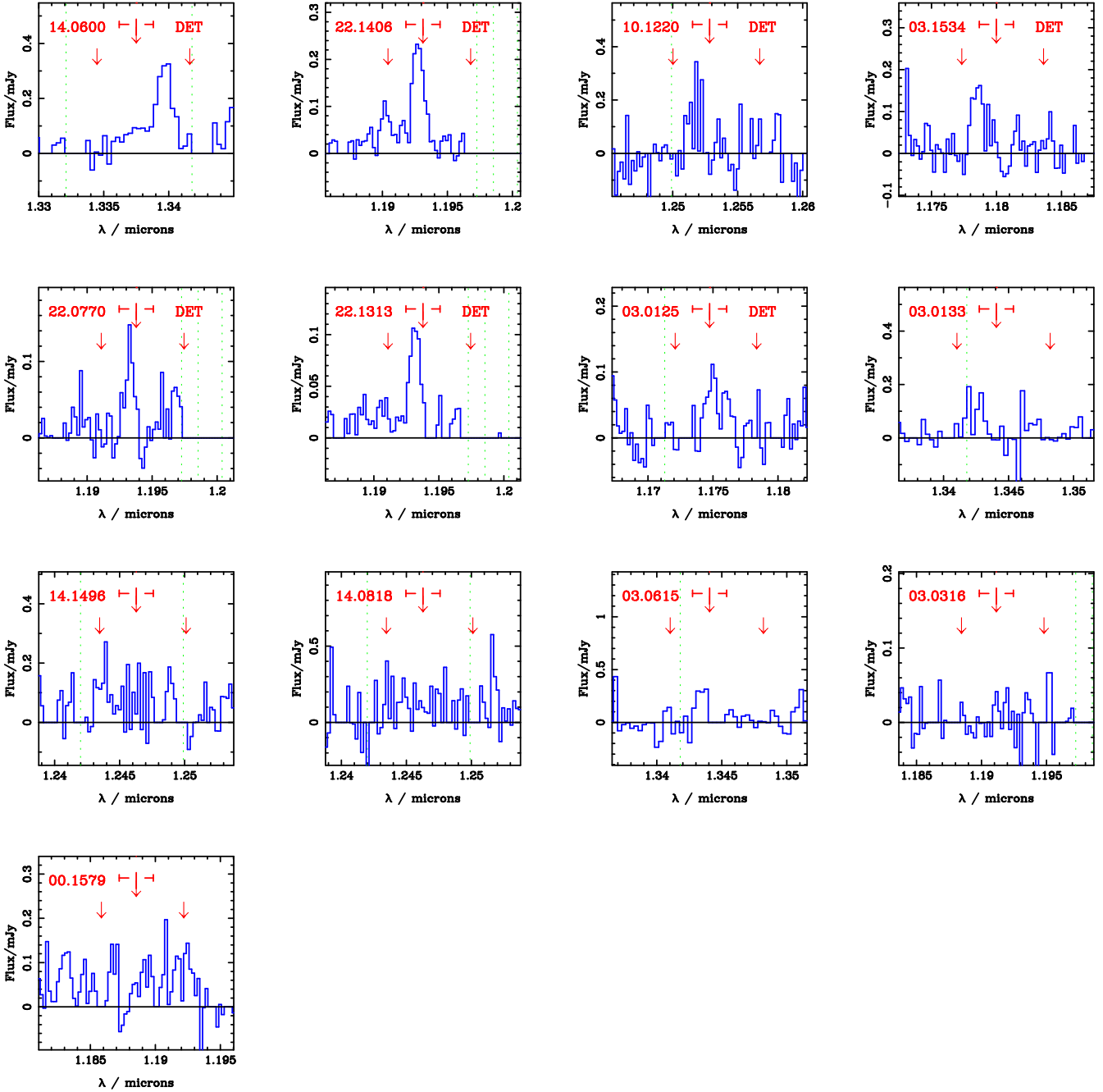}
\caption{\label{fig-spec}The $J$-band spectra of our CFRS galaxies. The spectra have
been continuum subtracted and centered on the \Halpha\ line. The large
central arrow indicates the predicted \Halpha\ position based on the
optical redshift, the associated horizontal error bar denoting the $1\sigma$
error on the redshift. The two smaller arrows show the predicted positions
of the [NII] 6548\AA\ and 6583\AA\ lines. The vertical dotted lines
indicate the positions of strong night sky OH lines which have been
masked out. The objects where we find a $>2\sigma$
\Halpha\ detection are labelled `DET.' }
\end{figure*}

Even with the resolved OH background accurate sky-subtraction is
critical to detection of faint lines. To first order the sky can be
removed by simply subtracted the pairs of offset frames, though this
leaves residual signal due to temporal sky changes. To second order
the residual sky was removed by performing a polynomial interpolation
along the slit, excluding the two object regions, and
subtracting. This leaves no systematic residual, though the regions
near the OH lines are still noisier due to the extra Poisson
contribution, with the result being a 2D image with a positive
spectrum in the `OBJECT' row and a negative spectrum in the `SKY' row.
For each image we also made a pixel mask to exclude bad pixels on the
detector and regions with noisy OH residuals.

In many of the images there were strong \Halpha\ and continuum
detections.  We summed all the \Halpha\ lines with good detections,
fitted Gaussians spectrally and spatially
to define the typical line profile and then used this 
mean profile (with the pixel mask) to
optimally extract all bright, faint and possibly non-detected objects
in a consistent manner.

In most of our observations we found that the negative spectrum was
typically weaker, or even absent, compared to the positive
spectrum. This is attributed to the fact that any small errors in
acquisition are magnified when stepping 9 arcsec away from the centre
along the slit as the rotation is not precisely known. When there was
a significant negative spectrum it was combined in a weighted manner
(to maintain consistent exposure times) into the positive
spectrum. This marginally increases the signal$/$noise, though none of
the flux measurements presented below are significantly changed if
this step is omitted.

Next the spectra had atmospheric absorption removed using a
smooth-spectrum standard and were flux calibrated using flux
standards.

Finally we applied aperture corrections to allow for our finite 1.0 arcsec
wide slit. Our objects are resolved, our mean Gaussian spatial profile
along the slit of the \Halpha\ line has a FWHM of 2.0 arcsec (which is
consistent with the typical 3--5 arcsec isophotal optical diameters
measured for the CFRS sample (Hammer \etal\ 1997)). The flux
calibration stars are observed through the same slit in 1.0 arcsec
seeing.  Assuming Gaussian profiles for both we derive a relative
correction factor of 1.7, by which we multiply our spectra. While this
can only be a rough correction, it gives fluxes consistent with the
optical lines (see section
\ref{sec-completeness}). It is also only of order unity, so even if ignored our
conclusions below are not drastically altered.

The continuum level was determined in each case by taking the mean of
the data points $\pm$ 0.02 microns either side of the emission line,
ignoring the masked points and the emission line.
The noise level of the data was determined empirically from the RMS in 
the same region. In most cases we found significant continuum emission.
Note that for individual pixels the continuum is mostly below the
noise, it is only by summing up we get a significant detection. To check
the validity of our measurements we repeated the same procedure for
an off-object row in the longslit spectra. This give non-detections
in all cases, so we are confident in our procedure.

After subtraction of the continuum level from the data we computed the line
flux by summing the flux $\pm N$ pixels around the line, excluding the
masked pixels, where $N=7$ is chosen to be the typical line FWHM. This
is done regardless of whether there appears to be a detection or not
(as the expected wavelength is known).

The final line fluxes and continuum fluxes (converted to AB mags and
denoted $J_C$) are given in Table~\ref{tab-observations}, along with
the most useful CFRS parameters of our objects. \Halpha\ fluxes
$<1\sigma$ are set $=0$.  We find 7 detections $>2 \sigma$ and 5
detections $>3 \sigma$ out of our 13 objects.

We convert these to luminosities using a $H_0=50$ km s$^{-1}$
Mpc,$^{-1}$ and $q_0=0.5$ cosmology in
Table~\ref{tab-luminosities}. We use this cosmology for the rest of
our paper. We note that in our further analysis our conclusions are
based on comparing the luminosities of the same galaxies at different
wavelengths, since all the galaxies in our sample lie close to $z=1$
our conclusions are essentially unchanged if we use a different
cosmology.

To complete the quantities derived from the lines we fit Gaussians profiles
(excluding masked pixels)
to derive velocity line widths for all our detections. Each of the fits
was checked visually by plotting on top of the line; the instrumental resolution
was determined by fitting to unresolved night-sky OH lines in the region
of the galaxy line and the value was subtracted in quadrature from the
galaxy line widths. It should be noted
that all our lines, but one, are well-resolved as we expected given the
spectral resolution and the typical velocity widths of galaxies. The
spectra line
Full Width Half Maximum (FWHM) of the galaxies
range from 3--5 pixels (one pixel
$=$ 2--3 \AA\ depending on the spectrum) compared to 2.4--2.6 pixels
for the sky lines. The results are presented in Table~\ref{tab-luminosities};
we make no detailed analysis here, we just note that the typical velocity
FHWMs are in the range 200--400 km$/$sec expected for large
$L\sim L_*$ galaxies.

\section{COMPLETENESS}  \label{sec-completeness}

In $8/13$ of the galaxy spectra \Halpha\ emission was detected;
additionally 
$9/13$ of the galaxies had detected continuum emission.  
This proves quite useful: we
can check if our line non-detections are due to poor acquisition by
comparing the continuum level of our line detections and
non-detections. There are only two cases in which there is no line {\em
and} no continuum detection.  The general trend is that the average
continuum flux is {\em brighter} for the non-detections, this
translates in to an median $(I - J_C)_{AB}$ colour 1.1 mags {\em
redder}. This argues against the slit missing the object, in fact the trend
towards redder colours is precisely what is expected for 
non-line emitting objects.  We note that at $z=1$ a Scd galaxy should
have observed colours $(I - J)_{AB}=1.0$ and an E/SO galaxy $(I -
J)_{AB}=2.0$, using template SEDs from Kennicutt (1992). We find
median colours of $(I - J)_{AB}=1.6$ for the \Halpha\ detections
and  $(I - J)_{AB}=2.7$ for the non-detections, which agree very well
given our actual galaxies will differ in detail from Kennicutt's templates.

Figure~\ref{fig-spec} shows the final set of spectra, which have been
continuum subtracted and centered on the \Halpha\ line. The line is in
all cases found within the error box given by the optical redshift
($\Delta z \simeq 0.002$).  Our resolution is high enough that
\Halpha\ is well separated from the [NII] lines so we do not have to
correct for blending. Also in a few cases (e.g objects 22.1406 and
22.070) we see evidence for one of the weaker [NII] lines as well as
the main \Halpha\ line. This is additional evidence for the robustness
of our detections.  Note in many cases the position of one or both of
the [NII] lines is occluded by a noisy OH residual.

\begin{figure} 
\psfig{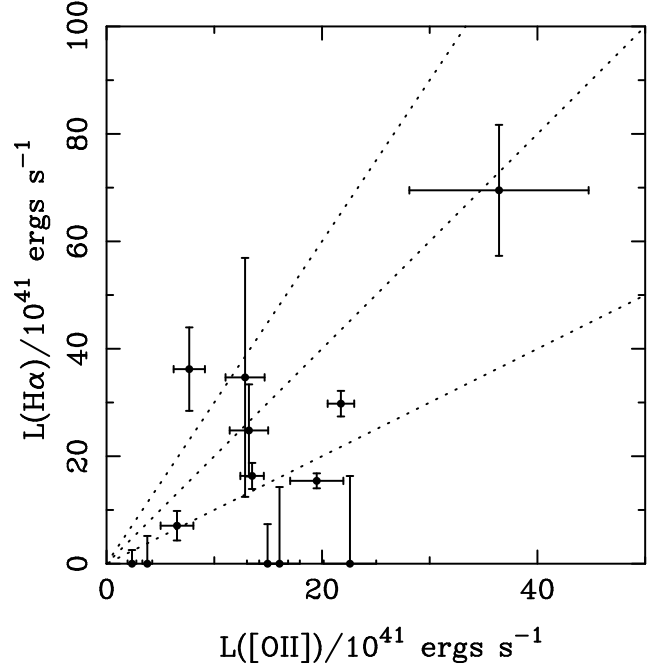}
\caption{\label{fig-oiiha}Comparison of line luminosities of the sample in [OII]
and \Halpha. Dotted lines show slopes of $\Halpha/\rm [OII]=1,2,3$, the spread
of ratios found be Kennicutt (1992).}
\end{figure}

We can also test our completeness using CFRS values for the [OII] flux
(Hammer \etal\ 1997) to calculate a \Halpha/[OII] line ratio. In the
CFRS sample in this redshift range 85\% of galaxies have [OII]
emission. Hammer \etal\ tabulate the [OII] equivalent width and
flux, the latter is aperture-corrected by comparing
the spectra at $\sim 5500\rm\AA$ with their $V$-band image photometry.
Other than excluding galaxies with zero [OII] emission we
made no attempt to concentrate on objects with the strongest [OII]
emission. Thus the mean [OII] equivalent width (32\AA) and range
(10--60\AA) of our small sub-sample are consistent with random
sampling from the larger sample.  The CFRS galaxies at $z=1$ with zero
[OII] are all at the red end of the colour distributions in $V-I$ and
$I-K$. Thus the colours also indicate they are not significant
star-forming systems and we conclude their exclusion has no
significant impact on our conclusions below.

One might also ask the question: is the lack of \Halpha\ detection in
some of our objects consistent with the {\em presence} of [OII] in our
optical spectra? This is addressed in Figure \ref{fig-oiiha} where we
plot the strength of the two lines against each other.  It can be seen
that given the \Halpha\ error bars all points are consistent with a
reasonable linear correlation. Kennicutt (1992) estimates the line
ratio $\rm\Halpha/[OII]$ as having
a median value of about 2 (with a spread from 1 to 3) for a sample
of local star-forming galaxies (1.0 mag mean
extinction). Our observed values at high-redshift
are entirely consistent with Kennicutt's median and spread, implying
we too have small amounts of extinction which agrees with our
findings below in 
Section~\ref{sec-sfrz1}. The points with zero \Halpha\ are
consistent with detected [OII] given the larger error bars.
Note the line ratios are also good evidence
that our aperture corrections are reasonable, if omitted we would
obtain a much lower ratio $\rm\Halpha/[OII]=1$.

\section{STAR FORMATION RATES FROM \Halpha\ and UV} \label{sec-sfrmethods}

Using models of population synthesis it is possible to calculate the
relation between input star-formation rates and output UV and \Halpha\
fluxes. The basic principle is that the UV light is dominated by
short-lived main sequence stars (the \Halpha\ light is reprocessed
ionising UV) so the number of them in a galaxy is proportional to the
star-formation rate. The prescription for this calculation is simple.
For reference we outline it (and the corresponding assumptions) in
detail:

\begin{enumerate}

\item For a given time-dependent star-formation rate the population
synthesis code gives the UV stellar spectrum as a function of time.
Note that Kennicutt (1983) in deriving his calibration uses a simple
grid of stars of different masses with evolutionary tracks current at
the time.

\item For the UV continuum estimators one takes an averaged flux (e.g. through
a synthetic box filter) at a specific wavelength (e.g. 1500\AA,
2800\AA). At the longer wavelengths, with increasing stellar
lifetimes, the conversion from a constant star-formation rate to a UV
flux is age-dependent (see for example Pettini \etal\ who uses
different factors at 1500\AA\ for $10^7$ and $10^9$ years).  One then
also needs to apply corrections for dust attenuation and/or Lyman
absorption by line of sight systems (e.g. Madau 1995).

\begin{figure} 
\psfig{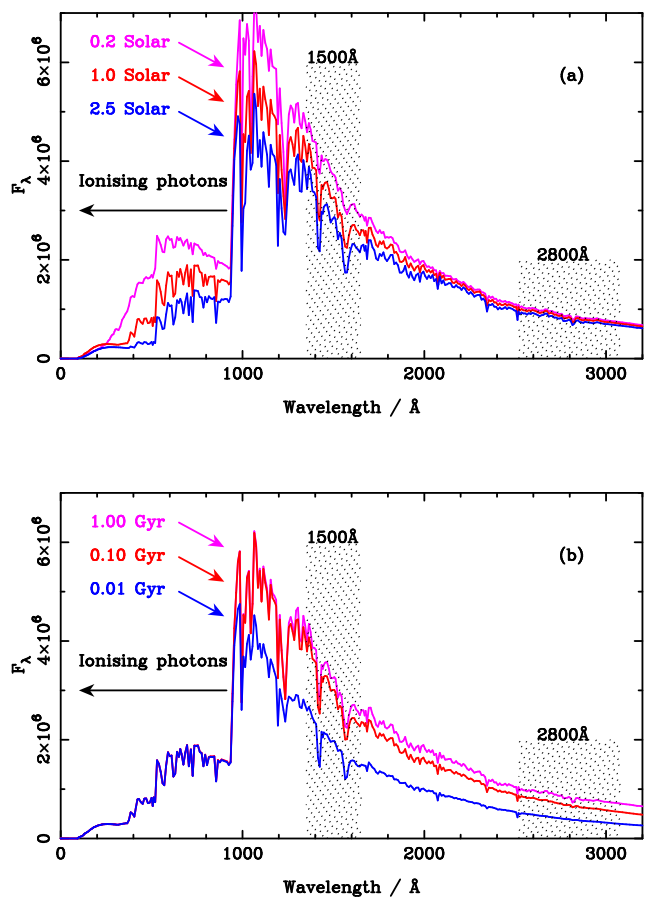}
\caption{\label{fig-uvspec} Illustration of the dependence of the UV
spectrum on metallicity and time for a constant star-formation
rate. These are theoretical galaxy spectra from the BC96 population
synthesis models (Salpeter IMF). The position of the 1500\AA\ and
2800\AA\ $\pm10\%$ bands are marked along with the ionising photons
which are reprocessed into \Halpha.  The vertical order of the
labels is the same as the vertical order of the spectra.
Panel (a) shows the spectrum at 1
Gyr for a decreasing metallicity sequence. Panel (b) shows the solar
metallicity spectrum for an increasing time sequence. It is easily
seen that while the redder UV bands are relatively insensitive to
metallicity they are strongly time dependent until 1 Gyr old. The
reverse is true for the ionising photons --- the stellar spectra 
below 912\AA\ are identical for times $>20$ Myr at a given
metallicity.  Hence the resulting \Halpha\ flux is mainly dependent
on metallicty.}

\end{figure}

\item For \Halpha\ one calculates the number of ionising Lyman
continuum photons. This flux comes from the most short lived stars
radiating at $\lambda<912$\AA\ with lifetimes of $\ls 10$ Myr. Then it
is assumed all this radiation is absorbed by intervening hydrogen gas
in the galaxy in which the forming stars are embedded and that none
leaks out.  In practice it seems this is very close to the truth.  At
very high-redshift the Lyman limit can be observed in the optical
(e.g. the Steidel \etal\ galaxies) and the flux does indeed go to
zero, but as these galaxies are identified by the Lyman breaks this
could be a selection effect. For local galaxies there have been
limited $\lambda<912$\AA\ spectroscopy, for example Leitherer \etal\
(1995) observed a sample of 4 starburst galaxies with the Hopkins
Ultraviolet Telescope and concluded that $<3\%$ of the ionising
photons escaped.  Constrained models of the ionising radiation field
of the Milky Way (Bland-Hawthorn \& Maloney 1997, 1998) indicate that
approximately 5\% may escape.

\item The ionising photons are reprocessed into recombination
lines, and the relative strengths can be calculated in detail.  Hummer
\& Storey (1987) calculate 0.45 \Halpha\ photons are emitted per Lyman
continuum photon for case B recombination.  This number is quite
robust, over a range of nebulosity conditions ($10^2-10^6$K,
$10^2-10^4$ electrons cm$^{-3}$).

\end{enumerate}

As there are a number of values for these conversion factors in the
literature we thought it would be useful for reference purposes to
systematically tabulate these for a set of models. This also serves to
illustrate the range of variations and trends. The results of our
calculations are given in Table~\ref{tab-conversions} for UV 1500\AA,
2800\AA\ and \Halpha\ conversions.  The Bruzual \& Charlot (1993,
1996) models (`BC96') offer a range of metallicities (albeit using
theoretical model atmospheres --- the `kl96' models).  The PEGASE
models (Fioc \etal\ 1997) offer two sets of post-main sequence
evolutionary tracks but only solar metallicity. Both sets of models
offer several Stellar Initial Mass Functions (IMFs); we tabulate the
results using the IMFs of Salpeter (1955) and Scalo (1986) because
these bracket the upper and lower limits of the variation.  We also
tabulate some other values given previously in the literature.

For all these models \Halpha\ reaches a constant asymptote after 20
Myr, then stays constant to within a 5--10\% out to 10 Gyr. Thus it
can be used as a tracer of the instantaneous star-formation rate on
time-scales of order $\sim$10 Myr. For reference we give the value at
100 Myr. The conversion is very sensitive to the IMF, as the ionising
flux comes from the most massive stars, with Salpeter giving a value
2.8--3.4 $\times$ higher than Scalo for the different models. Another
important point, which has not been remarked upon in the literature
discussing high-$z$ galaxies,
is the strong effect of metallicity below 912\AA. This occurs
in the stellar evolutionary models
because the most luminous stars (contributing below 912\AA), have 
a much greater dependence of effective temperature on metallicity 
than the less luminous stars contributing at longer wavelengths. 
The effect of the hotter stars at low metallicty is to raise 
the \Halpha\ flux about 1.7 $\times$ compared to the solar
metallicity value for a given star-formation rate.

The average UV luminosity per unit Hz is computed as the total energy
integrated through a $\pm$ 10\% box filter divided by the total
bandwidth. This gives a negligible colour term. For a 10\% filter the
difference in the average between a constant $f_\nu$ spectrum (which
approximates these spectra for $>912\AA$) and a much redder constant
$f_\lambda$ spectrum is only 1\%. The IMF dependence is somewhat less
than for \Halpha\ (typically a factor of 1.5--2.5) as these UV bands
come from less massive stars and the metallicity dependence on
temperature is much smaller.

However there is a stronger time dependence, with the flux increasing
significantly between $10^7$ and $10^{10}$ years, especially at
2800\AA\ and for the Scalo IMF which is much less rich in massive
stars. For cosmological times ($>1$ Gyr) the 1500\AA\ flux is
relatively insensitive to time as noted by Madau \etal\ (1998) (except
for the extreme 0.02 Solar metallicity case). Thus it can be used as a
tracer of the star-formation rate on time-scales of 1 Gyr. The
2800\AA\ flux is not as stable and changes at the 10--40\% level over
1--10 Gyr. There are further $\ls 50\%$ variations in the exact
conversion depending on the model and metallicity assumed.

These time and metallicity effects are illustrated graphically in 
Figure~\ref{fig-uvspec}. Finally it is worth noting that the 
range in the {\em ratio} of $\Halpha$ to UV is less than
the absolute range; as both come from high-mass stars the choice
of IMF matters somewhat less.

\section{THE STAR-FORMATION RATE AT $z=1$} \label{sec-sfrz1}

Turning back to our data we can do a direct galaxy by
galaxy comparison of the star-formation rates inferred
from UV and \Halpha. 

Our UV fluxes are particularly robust because for redshifts near unity
the rest frame 2800\AA\ light corresponds very closely to the observed
frame $V$-band light. To a first approximation we can simply ignore
the K-correction and derive the 2800\AA\ flux directly from the CFRS
$V$-magnitudes. We refine this slightly by using the SED fits from
Lilly \etal\ (1996); this corrects the fluxes by 10--20\%. Thus the UV
luminosities we use are the same as the raw data which goes into the
$z=1$ star-formation rate determinations of Madau \etal\
(for $0.2\le z\le 1$ this was based on the CFRS luminosity density
functions of Lilly \etal\ 1996)
We use a formal error bar of 20\% for our UV fluxes to give
errors representative of the $V$ photometric errors and the
K-corrections.

\begin{figure} 
\psfig{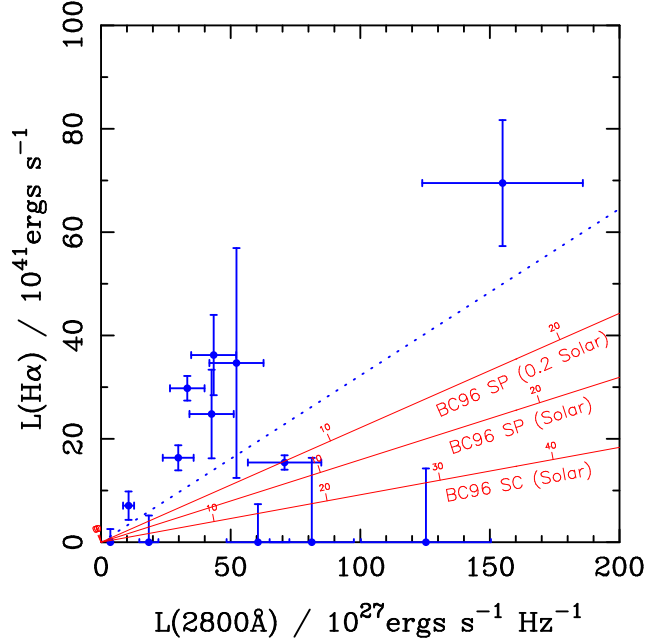}
\caption{\label{fig-havsuv}Comparison of the \Halpha\ vs UV continuum
flux at 2800\AA\ for the individual galaxies. The overlayed axes
show the locus of constant star-formation rate for a set of fiducial models
covering the range of \HUV\ ratios given in the last column of
Table~\ref{tab-conversions}. The numeric labels on the axes are the
corresponding star-formation rates in \Msun\,yr$^{-1}$. The dotted line
shows the ratio of the mean luminosity densities derived in 
Section~\ref{sec-sfrz1}.}
\end{figure}

We plot the \Halpha\ vs UV luminosities in Figure~\ref{fig-havsuv} and
overlay lines for a set of conversions from
Table~\ref{tab-conversions}.  Note these are only valid for a
continuous, constant star-formation rate --- the more complex problem
of bursts is considered below in Section~\ref{sec-models}.

It is clear from the plot that there is a order-of-magnitude
agreement in the star-formation rates derived from
the two methods and a reasonable correlation, i.e. strong UV systems
are usually strong \Halpha\ systems. There are no galaxies with an
extremely large excess of \Halpha\ relative to UV, which would occur
if the star-formation was highly obscured by dust.

We do find points with zero \Halpha\ flux and appreciable UV flux,
which we would expect to detect within the measurement errors if there
was a linear correlation. This can not be due to dust as the latter
can only enhance the \HUV\ ratio not diminish it. As noted in
Section~\ref{sec-completeness}, these spectra are detected, they just
do not contain significant \Halpha.  One physical effect that will explain this is
the relative lifetimes of stars contributing to the UV and \Halpha\ as
mentioned in Section~\ref{sec-sfrmethods}. When one moves away from
the simple scenario of constant star-formation rate the picture
changes considerably. For example for a instantaneous starburst the
\Halpha\ flux will drop to effectively zero after 30 Myr while
significant UV flux will persist up to 1000 Myr.  This will produce
low \Halpha\ points such as are seen in our sample. {\em Weak\/} \Halpha\ should
be present in {\em all our galaxies} because they are known to have [OII]
emission (see Section~\ref{sec-completeness}), however the \HUV\
ratio will scatter more widely and it is certainly possible for 
\Halpha\ to be below our detection limit despite significant [OII] and
UV emission. To quantify these effects
requires us to develop a proper model for starburst activity in
galaxies. We do this below in Section~\ref{sec-models}.

Nevertheless when one averages over many galaxies the effect of bursts
should cancel out in the mean --- the continuous star-formation
conversions of Table~\ref{tab-conversions} should be applicable for an
{\em ensemble} of galaxies. (This approximation is tested below using
the methods developed in Section~\ref{sec-models} and is found to
accurate to $\sim \pm 10\%$, even for this small sample.)  It can be
seen that the values scatter about a ratio of 2--3 times as much star-formation
inferred from the \Halpha\ as UV. This is balanced to some extent by
the zero points. By summing over {\em all} the galaxies we find the
following relation between the luminosity per comoving volume in \Halpha\
and UV:
$${\displaystyle   L(2800\rm\AA) \over \displaystyle L(\Halpha)} = 3.1 \pm 0.4 $$
in the units as in the last column
of Table~\ref{tab-conversions}. This ratio is plotted as
the dotted line in Figure~\ref{fig-havsuv}. 

The ratio is somewhat lower that those predicted by the models which
give values in the range 4--14 for 0.2-2.5 Solar metallicities. What
can cause this discrepancy? It is too large to be encompassed by
our range of metallicities; one can boost the \HUV\ ratio
if we assume a younger age than $\sim 1$ Gyr. However changing to $0.1$
Gyr only increases the slopes of the solid lines in Figure~\ref{fig-havsuv}
by $\sim 30\%$; we could achieve the 
observed slope if star-formation is only $\sim 0.01$ Gyr old in all the
detected galaxies but this would be unlikely simply due to random sampling
--- the redshift range of the sample
corresponds to a timespan of $\simeq 1$ Gyr at $z=1$.
The most likely explanation for the discrepancy is the presence of
dust attenuating the ultraviolet light. 

If we adopt fiducial model values of (10.9, 6.3) for (Scalo, Salpeter) for
Solar metallicity, which agree well between PEGASE and BC96 models
(kl96 tracks), the ratio of
\Halpha\ to UV inferred star-formation rates are (3.5, 2.0).
Using Pei (1992) extinction formulae (a standard dust-screen model)
we then derive 
a mean sample $A_V$ of (1.0, 0.5)
mags for the SMC law, (1.1,0.6) for the Galactic law.  

These extinction values are entirely consistent with what is found from
studies of star-formation from Balmer lines in local normal Sab
galaxies.  For example Kennicutt (1983) found $A_V=1.0$ mags and a
study of low redshift $z<0.3$ CFRS galaxies using the $H\beta/\Halpha$
line ratios by Tresse and Maddox (1998) found the same value.

Calzetti \etal\
(1994) and Calzetti (1997) have proposed an empirical dust-attenuation law for
heavily reddened starbursts ($A_V \simeq 2.2$ mags). In this 
model the nebular lines have about twice the optical depth as the
stellar continuum; moreover although the lines are well described by
a standard Galactic screen law the stellar continuum is empirically described
by a greyer law in which the dust and stars are intermixed. 
Using the Calzetti law we derive
$A_V$ attenuations of (2.6, 1.4) mags (for stars; $\times 1.4$ for nebulae);
because the attenutations of the \Halpha\
and  2800\AA\ continuum are more similar than for the screen models a
much higher obscuration is required to match the observed excesses:
attenuations of (2.9, 1.6) mags result for \Halpha\ and (4.2, 2.3) mags
for 2800\AA\ stellar continuum. It should be noted however that it is
still not established that corrections derived for the Balmer lines
in starburst regions of nearby galaxies are appropriate to the integrated
light of the distant galaxies studied here. This question remains open.

We can now examine the total correction for dust in both the
\Halpha\ and UV determined star-formation rates. The Milky way law
gives corrections for (Scalo, Salpeter) IMFs of $\times(2.2,1.5)$ for \Halpha\
and (8.0,3.1) for the UV. These
numbers are the same to within $\pm 10\%$ for the SMC extinction law;
this is because the two extinction curves differ most strongly in the UV at
$<<2800\rm\AA$ (e.g. at the 2175\AA\ dust feature) and in the near-UV
and optical they are very similar. The Calzetti law of course gives
much larger values: the final star-formation rates are an additional factor of 
$\sim 6\times$ higher for the Scalo IMF and $\sim 3\times$ higher for 
the Salpeter IMF. 

\begin{figure*} 
\psfig{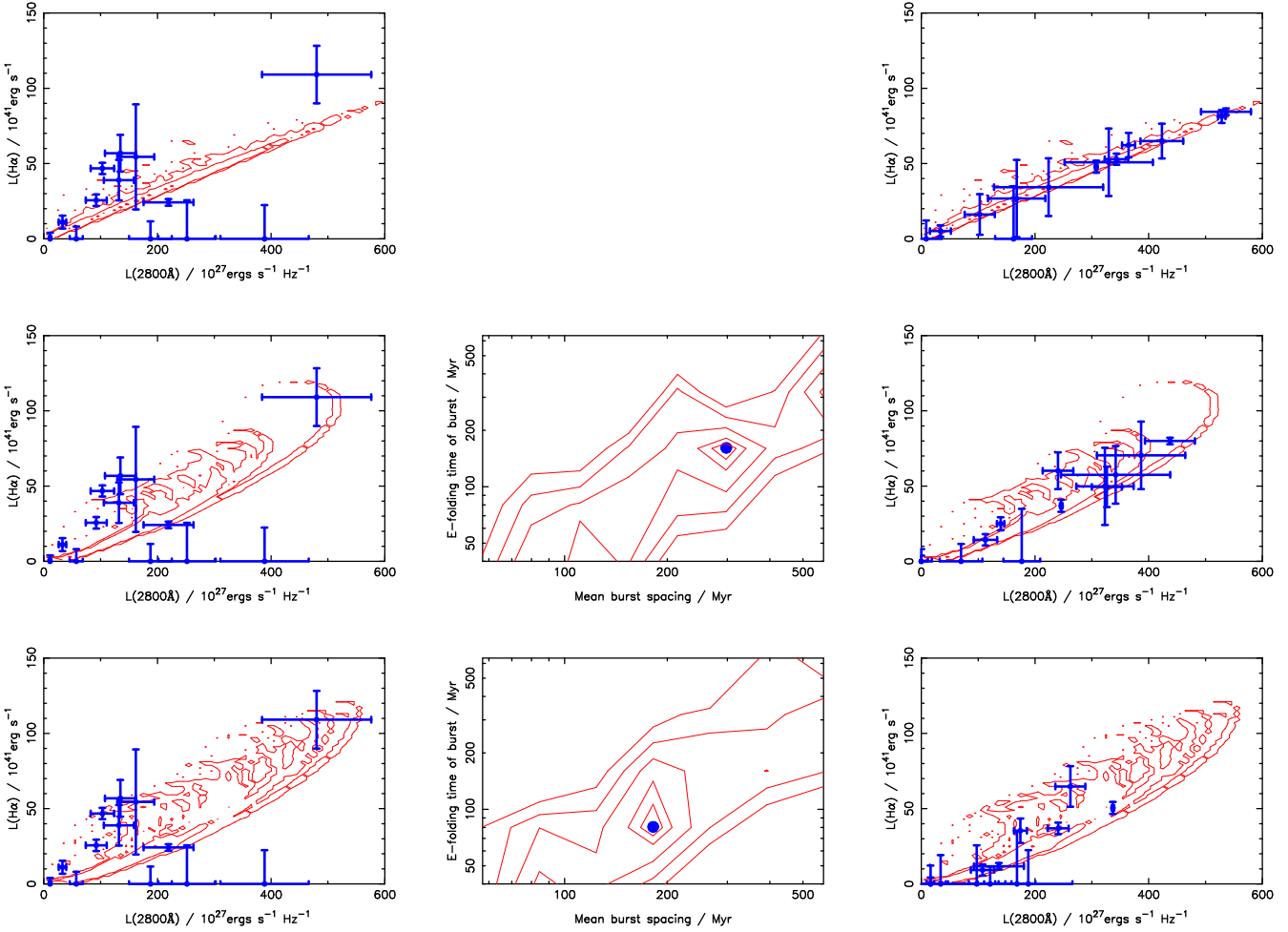}
\caption{\label{fig-modelbursts}Results of fitting burst models to
our extinction-corrected \Halpha\ and UV (2800\AA) data. The three
rows show three different models as described
in the text, they are: continuous star-formation (top row), fixed mass
exponential bursts plus continuous star-formation (middle row) and 
variable mass exponential bursts plus continuous star-formation (bottom row).
The left hand column of panels shows
the observational data (points) compared with a model distribution
generated from the best fit model (contours correspond to a factor of 10
in probability density). The middle column of panels show the
likelihood contours of the main burst parameters generated by our fitting. (Contours
correspond to a factor of 10
in likelihood, the
circle marks the maximum likelihood point.)
The right hand column show a {\em realisation} of the best fit model, {\it i.e.}
13 simulated data points generated from the model distribution with the
observed errors (points $<1\sigma$ are set $=0$ as in the data). 
Note the continuous star-formation includes a component of
galaxy--galaxy scatter.}
\end{figure*}

Finally we can compare the star-formation rate of our $z\sim 1$ CFRS galaxies
with local counterparts. We adopt the Salpeter IMF as that is
conventionally used for
deriving the local rates. For the range of dust
corrections we have derived we obtain rates of $\sim 20$--60 \Msun\ yr$^{-1}$,
comparable to local starbursts
(e.g. Calzetti 1997) and much greater than the typical 4 \Msun\
yr$^{-1}$ found for local normal spirals (Kennicutt 1983) and the
Milky Way (Smith \etal\ 1978).

\section{MODELING OF STARBURSTS} \label{sec-models}

As noted earlier the interpretation of \Halpha\ and UV
luminosities as star-formation rates becomes more complicated
if non-constant star-formation histories are assumed. For
this reason we developed a mathematical frame-work for exploring this
and to see how well we could reproduce the observed
distribution.

\subsection{Methods}

The principles are based upon maximum likelihood and are a 2D
generalisation of the methods developed by Abraham \etal\ 1999 for
colour-colour fitting. For a given star-formation history we can run a
spectral evolution code and calculate \Halpha\ and UV luminosity as a
function of time using the methods of Section~\ref{sec-sfrmethods}. We
evolve the models for 5 Gyr and sample at random times (we can do this
because all the galaxies are at similar redshift) to create a 2D model
distribution of light in the the $\rm(\Halpha,UV)$ plane.  The code
then computes the likelihood of drawing the observational data from
the model:

\def\minfty{\kern -2.5ex\lower 1ex\hbox{${}_{-\infty}$}}
\def\pinfty{\kern -1ex\raise 0,5ex\hbox{${}^{\infty}$}}
$$ {\cal L} = \prod_i \int^{\pinfty}_{\minfty}\int^{\pinfty}_{\minfty}
\raise-1em\hbox{${\displaystyle P(h,u) \exp\left(
- {(h_i - h)^2 \over 2\Delta h_i^2} - {(u_i - u)^2 \over 2\Delta u_i^2}
\right)\over \displaystyle \vphantom{\sum} 2\pi \Delta h_i \Delta u_i}$} du \,dh
$$

where the observational $\rm(\Halpha, UV)$ points are $(h_i, u_i)$ with
errors $(\Delta h_i,\Delta u_i)$,
and the model points have probability density $P(h,u)$.

The star-formation histories are parameterised and likelihood space
is explored to find the maximum likelihood. 
We use an adaptive algorithm where a large parameter
space is explored with a coarse parameter grid to locate the peak
and the region around the peak is then examined with a finer grid
to give confidence limits on the fitted parameters from $\Delta {\cal L}$.

Once we have the best fit parameters for a model, we can then
create simulated observational datasets. This is done 1000 times,
the maximum likelihood fit being recomputed each time, this allows us to
normalise our relative likelihood into an absolute probability
of the observed data given the model.

\subsection{Star-formation histories explored}

\begin{figure} 
\psfig{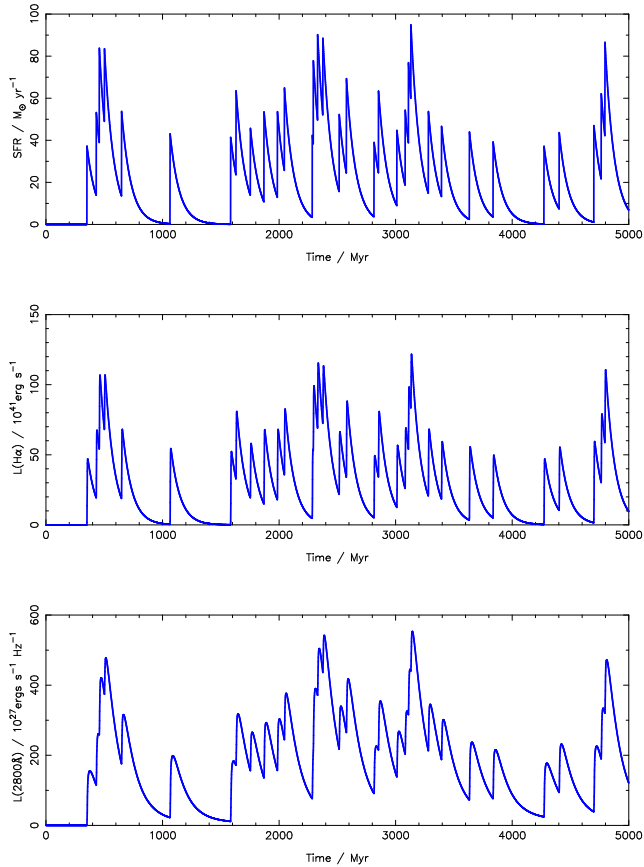}
\caption{\label{fig-modeltime}The star-formation rate,
luminosity history in \Halpha\ and UV (2800\AA)
typical of our best fit models. This is for the case of exponential bursts
of variable mass, with a continuous star-formation component, {\it i.e.}
the model shown in the lower 3 panels in Figure~\ref{fig-modelbursts}. It can
be seen that for long inter-burst periods galaxies 
are indeed quiescent in \Halpha,
but not the UV.
}
\end{figure}

We parameterised the star-formation histories as continuous star-formation,
with galaxy to galaxy scatter, plus a random distribution of bursts. Initially
the total star-formation rate is kept constant and normalised to the
values derived in Section~\ref{sec-sfrz1}. This is a good approximation
as an ensemble of galaxies undergoing bursts will approximate
continuous star-formation and results in one less free parameter.
For our further analysis we confine ourselves to the BC96 models
(Solar metallicity, `kl96' atmospheres) and the Salpeter IMF
(as the latter gives better results for fitting the star-formation
histories, colours and mass$/$light ratios of galaxies, see for example
Kennicutt (1983) Madau \etal\ 1998,
Calzetti 1997, Lilly \etal\ 1996). We correct all
$\rm(\Halpha, UV)$ points using the
$A_V=0.6$ mags Milky Way law derived in Section~\ref{sec-sfrz1}; the
effect of adopting the Calzetti law is just a simple scaling to globally 
$3\times$ higher star-formation rates and does not affect the details of the analysis.

To start with we modeled a simple continuous star-formation model
to check our code was giving sensible results. To make it more
realistic we introduced a scatter, $C_\sigma$, between galaxies
following a normal distribution (slightly corrected to avoid
negative star-formation rates).

The best fit results are shown in Figure~\ref{fig-modelbursts} as a likelihood
contour plot overlayed on the model points. For comparison we also
show a simulated set of data points drawn from the model distribution.
It can be seen that some scatter is introduced;  this originates
physically from the time variation of UV light even for constant
star-formation. The best fit parameter values are shown in
Table~\ref{tab-likepars} --- the model is a very bad fit to the
distribution of data.

One might ask if the scatter could be explained by variation in
extinction between galaxies. While it is possible this can explain
some of the variation it can not explain the points with large
amounts of UV emission and small amounts of \Halpha\ emission.
This is because dust quenches the UV much more than the \Halpha\ ---
precisely the opposite effect to that sought. We next explored
the effects of starbursts to see if these could plausibly
explain the observed distribution.

Initially we tried fixed-mass
bursts, then we tried a scheme for allowing the burst
masses to vary. Since we could find little
information in the literature as to an appropriate mass function
to adopt for bursts we invented our own simple phenomenological scheme:
bursts are parameterised by a mean mass ($M_\mu$) and
standard deviation ($M_\sigma$). The distribution is assumed
to be normal. While this has no physical basis it only has
2 parameters and at least allows us to investigate the effects
of mass distributions. $M_\mu$ and our global star-formation
rate normalisation fixes the mean interval between bursts.
Finally we have $F,$ the fraction of the star-formation
occurring in bursts. For the form of the bursts we consider
two cases (modeled after BC96): a constant burst of length $\tau$ and 
exponential bursts of e-folding time $\tau.$

The results of this exercise is shown in Table~\ref{tab-likepars}. It can
be seen that the burst models provide much better fits than the
the continuous models. The Monte-Carlo realisations show that the latter
generate data sets like the observed one only about 1\% of the time.
This is because the continuous star-formation can not generate enough
variation in the \HUV\ ratio. The burst models are much better
and generate synthetic points which look
like the data $\sim 10$--20\% of the time. This is because
the variation in the star-formation rate is the
principle cause of variation in the \HUV\
ratio. Allowing the burst mass to vary improves the fit only slightly, we
conclude that our data does not constrain the shape of
the burst mass function significantly.
The best fit burst fraction is $\simeq 1$, indicating the burst
mode is preferred. The results are illustrated graphically in the lower two
rows of
Figure~\ref{fig-modelbursts} which compares the dust-corrected
observational data
with model realisations for a sample of key models. The likelihood
contours of the main parameters are also shown.

The best fit mass of a typical burst is 2--5$\times 10^9$ \Msun, corresponding
to a time interval between bursts of typically $\sim 200$--300 Myr 
and the characteristic
time $\tau$ is of $\sim 100$--200 Myr. This mean the bursts usually overlap
in time. These values are similar to 
what one might expect intuitively based on the data: the chance of
catching a galaxy in the \Halpha\ quiescent stage has to be of order $1/3$
to reproduce the fraction of points seen with UV but no \Halpha. 

\begin{figure*}
\psfig{file=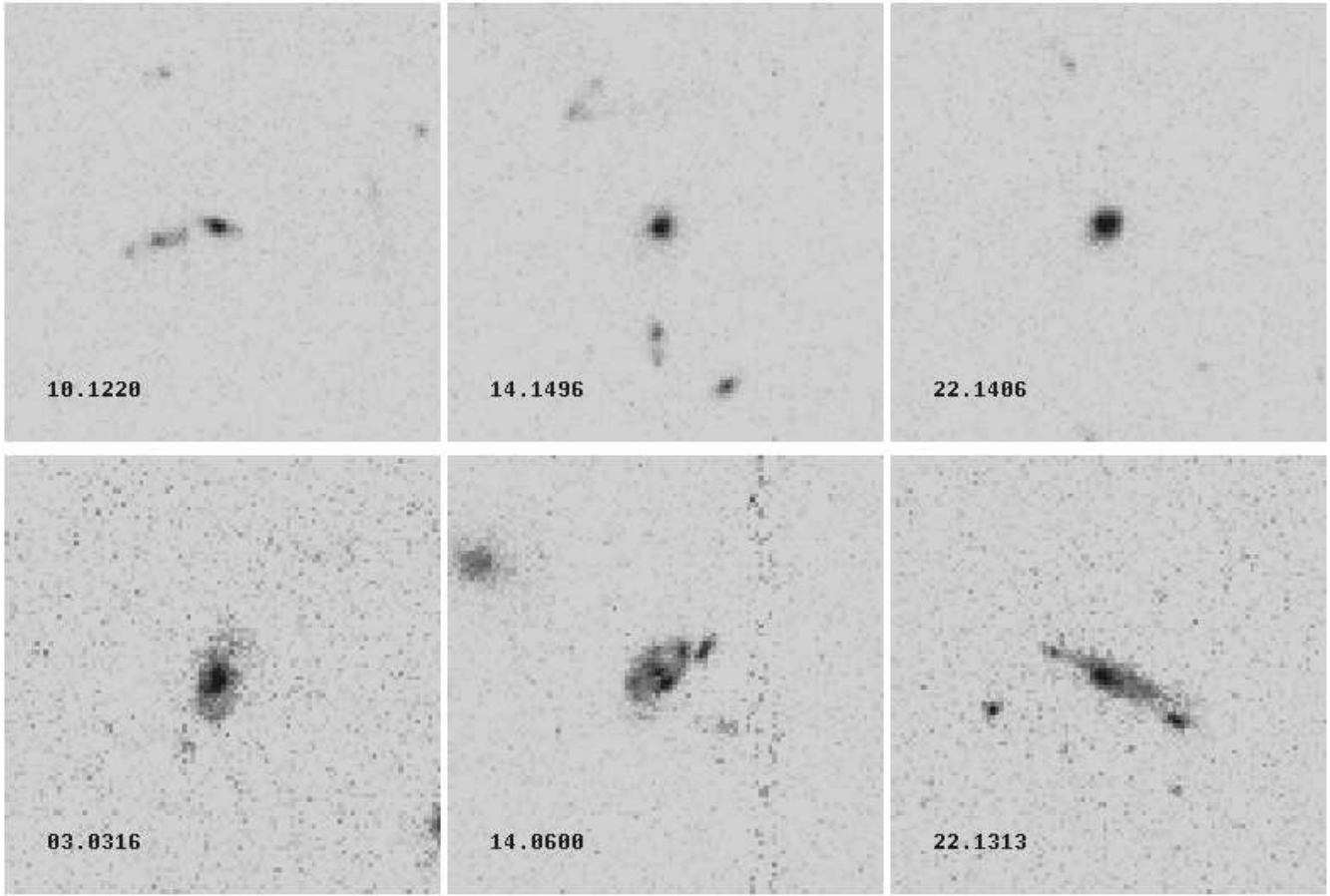,width=\hsize}
\caption{Postage stamps ($10.1 \times 10.1$ arcsec)
of the six galaxies having images from our
HST morphology programme.\label{fig-postage}}
\end{figure*}

We note that this kind of
`continuous but episodic' star-formation with several bursts per
Gyr is of the same form as that found in local starburst galaxies
by Calzetti (1997). Our average star-formation rates are of the 
same order too. 
In between bursts the star-formation rate and \Halpha\ flux does indeed drop
close to zero while the UV persists (see Figure~\ref{fig-modeltime}
which shows a sample time-dependence)
due to the stellar lifetime effects mentioned in Section~\ref{sec-sfrmethods}.
Thus the zero \Halpha\ points in our data (given the error bars)
are naturally explained. 

Finally with these tools we were able to test how well an ensemble of
galaxies converged to approximating a continuous star-formation
rate, as assumed in Section~\ref{sec-sfrz1}. To do this we re-ran the
likelihood fitting, this time fitting for the total star-formation rate
as a free parameter. The results of this gave rates
of between 19 and 23 \Msun\ yr$^{-1}$ per galaxy, which agrees well
with the value of 20  \Msun\ yr$^{-1}$ calculated in Section~\ref{sec-sfrz1}
for the same Salpeter IMF.

While these simple models could do with some elaboration to obtain a better fit,
we are near the limit of what can be inferred from 13 data points. It
is clear this sort of detailed approach will benefit greatly from future 
observations
and much larger samples.

\section{MORPHOLOGICAL TRENDS} \label{sec-morph}

Six of the galaxies in our sample have morphological information
from our programme of {\it Hubble Space Telescope}
high-resolution imaging of CFRS and LDSS2 high-redshift
galaxies (Brinchmann \etal 1998, Lilly \etal, 1998). This is
obviously an even more limited sample, but we can look
qualitatively at the dependence of star-formation rate on galaxy
type.

\begin{figure*} 
\psfig{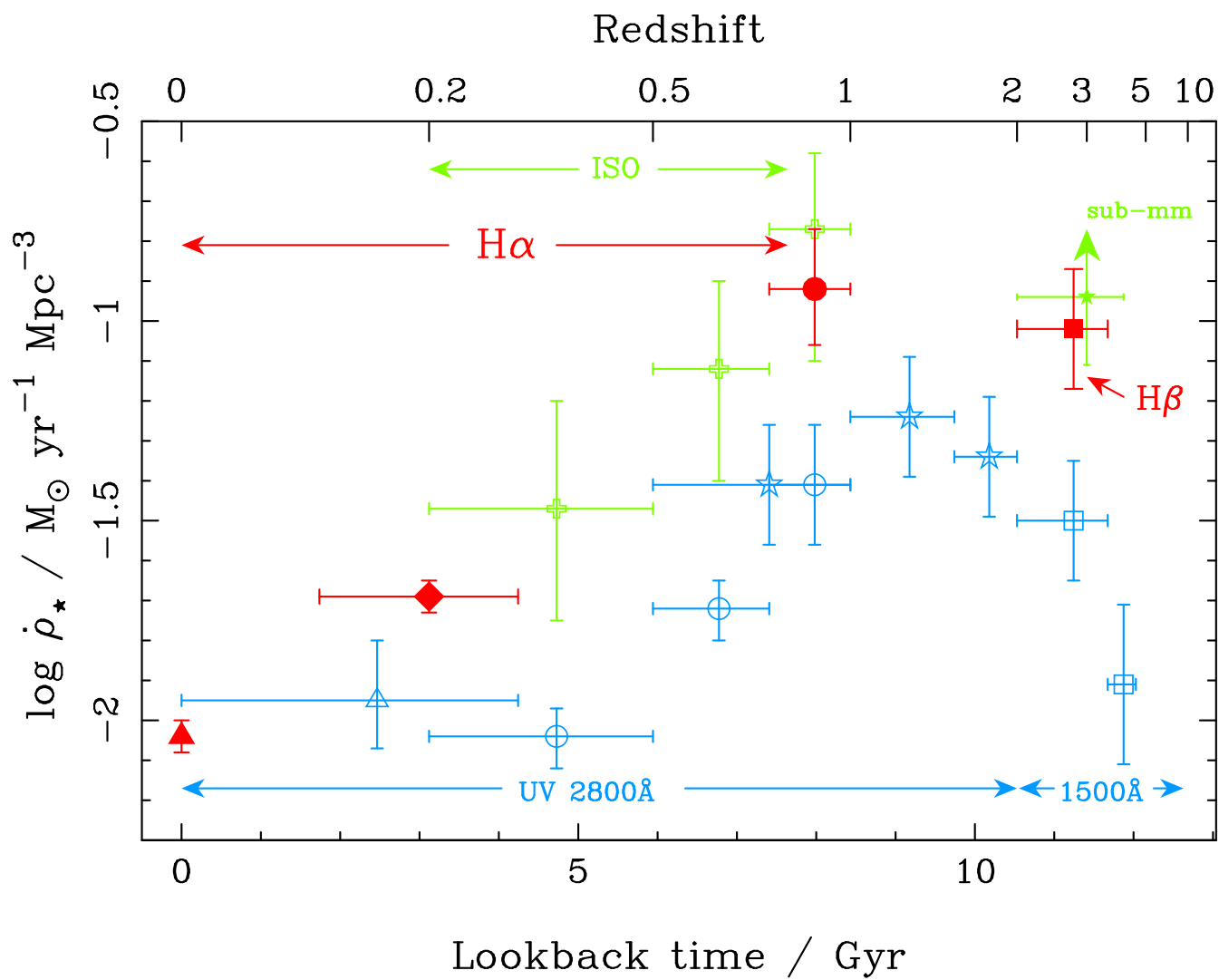}
\caption{\label{fig-balmer-sfrz}Star-formation history of the Universe
inferred from Balmer lines (\Halpha\ and \Hbeta) 
compared to UV and far-IR/sub-mm determinations. The UV points
are: Treyer \etal\ (1998; open triangle), Lilly \etal\ (1996; open circles),
Connolly \etal\ (1997; open stars), Madau \etal\ (1996, 1998; open squares).
The Balmer points are Gallego \etal\ (1995; filled triangle), Tresse \&
Maddox (1998; filled diamond), this work (filled circle), Pettini
\etal\ (1998) correction of the UV Madau \etal\ point at $z=2.8$
(filled square). 
We also plot the sub-mm derived point of Hughes \etal\ 1998 (solid star) and the 
far-IR ISO derived
points of Flores \etal\ (1998) (open cross). 
}
\end{figure*}

The classifications are listed in Table~\ref{tab-observations} and
postage stamps of the galaxies are shown in
Figure~\ref{fig-postage}. There are 3 galaxies classified as
`Peculiar'. All have detected \Halpha\ emission and blue colours 
($(V-I)_{AB}<1$) and one (14.0600) has the highest star-formation rate in our
sample. Two of these are classed as `mergers' and one as a close pair
indicating an association between star-formation and
interaction. There are two galaxies classed as 'Compact'. Both of
these are also blue ($(V-I)_{AB}<1.2$) and it is interesting to note
that while one has quiescent \Halpha\ and strong UV the other has a
large \Halpha\ excess (3.2). Finally we have 03.0316 a red spiral
($(V-I)_{AB}=2.9$) which is quiescent in both UV and \Halpha.

Finally we note that galaxies with star-formation, whether
inferred through \Halpha\ or UV, have the bluest $V-I$ colours
($(V-I)_{AB}<1.2$).  This is not in contradiction with the modest
extinction values derived in Section~\ref{sec-sfrz1} --- an extinction
of $A_V=0.6$ mags would only redden the observed $V-I$ (rest $2800\rm\AA-B$)
by a small 0.4 mags whereas
the difference between and old and young stellar population is of
order 2--3 mags.

\section{COMPARISON WITH OTHER RESULTS} \label{sec-comparison}

There are now enough measurements of Balmer line star-formation rates
at high and low redshift to construct the first star-formation history
of the Universe in Balmer light to compare with the previous UV
measurements. For consistency we use the Salpeter IMF, $H_0=50$
\Hunits, $q_0=0.5$ throughout. All points are re-derived from their
original luminosity densities in a consistent manner using the UV,
\Halpha\ factors in Table~\ref{tab-conversions} for BC96 (kl96) with
Solar metallicity.

This is shown in Figure~\ref{fig-balmer-sfrz}. The point at $z=0$
comes from the Gallego \etal\ (1995) local objective prism survey and
is based on \Halpha. Tresse and Maddox (1998) have measured the \Halpha\
luminosity function at $z=0.2$ from the CFRS, at which point \Halpha\
is still available in the optical CFRS spectra. They find a value
a factor of two higher than the UV measurements.

Our \Halpha\ measurements are used to derive a new value for the
star-formation rate density in the CFRS at $z=1$. This is higher
by a factor of 3.1$\times$ than the UV point. 
At $z=2.8$ we show the point derived
from the work of Pettini \etal, who used CGS4 to measure
the \Hbeta\ line in 5 of Steidel \etal's galaxies. They infer star-formation
rates 0.7-7 times higher than derived from the UV at 1500\AA\ rest
and typical extinctions $A(1500\rm\AA) = 1$--2 mags. We show this as 
a factor of 3 above the UV point.
As well as the Balmer lines 
we also plot the point of Hughes \etal\ based on sub-mm observations and the
points of Flores \etal\ (1998) from far-infrared ISO observations. It
should be noted though that the derivation of the latter are qualitatively different
from the Balmer line and UV measurements:
the far-IR and sub-mm bands measure UV reprocessed by dust 
into thermal radiation and hence they are
sensitive to galaxies which might not appear at all in the optical. Moreover
the Hughes \etal\ points
are based on an {\em assumed\/} redshift distributions for sources
which have not yet been verified and has been disputed
(Richards 1999). Finally it is also worth noting that the
Hughes \etal, Connolly \etal\ and  Madau \etal\ points
are all based on the same patch of sky, the Hubble Deep Field North (Williams \etal\
1996), which may not be representative.

It can be however seen that the general trend is for the Balmer line and
ISO/sub-mm measurements to find values several times higher than the
UV continuum at all redshifts. 
The rise to $z=1$ is preserved, arguably the fall off at
$z>2$ is preserved though given the random errors and the systematics
in the dust correction no change for $z>1$ would also be 
consistent with the data. Whether star-formation peaks at intermediate
redshift ($z=1$--2) or continues to high redshifts ($z>4$) is an
important test of heirarchical formation scenarios (e.g. Baugh \etal\ 1998).
From the current data the question must remain open.

The agreement between the far-IR/sub-mm measurements
and Balmer measurements is particularly impressive since both
attempt to compensate for dust in different ways. Flores \etal\
find an upward dust correction of $2.9\pm 1.3$ at $z=1$ and extinctions
of $A_V=0.5$--0.9 mags, both consistent with our best estimates. It should
be noted that independent ISO observations of the Hubble Deep Field North
by Rowan-Robinson \etal\ (1997) give a conflicting value several times
higher than Flores \etal; however the latter is derived from
a 19$\times$ larger area of sky and so is probably better determined.

An important point is that the dust corrections are not many times {\em larger}
than we have found. Larger corrections (e.g. $\times 15$) have been 
argued for by authors such as
Meurer \etal\ (1997) based on amounts of obscuration in powerful
starburst galaxies locally.  However it is not clear that the local
massive starburst galaxies (e.g. Arp 220) are the right analogs of the
higher redshift field galaxies. When lower luminosity starburst galaxies
are studied locally they prove much less obscured (e.g. Buat \& 
Burgarella 1998), these systems are comparable in extinction 
to local spirals and the sample studied here. We note though if
we use the dust correction from the Calzetti law the \Halpha\
point at $z=1$ moves up still further to $9\times$ above the UV
point. This is because if our galaxies are similar to those
studied by Calzetti the HII regions suffer additional obscuration
relative to the stellar continuum. This higher value would then agree with
the ISO observations of Rowan-Robinson \etal. However the resulting
high star-formation rates may be problematic in that the stellar density
produced in the local universe might be too high (Madau \etal\ 1998).

\section{CONCLUSIONS} \label{sec-conclusions}

From a sample of 13 galaxies observed with 
CGS4 we have performed the first \Halpha\ measurements
of the star-formation rate at $z=1$. We conclude
the following:

\begin{enumerate}

\item The \Halpha\ measurements show a star-formation rate at least
three times as high as that inferred from the 2800\AA\ continuum luminosity
by Madau \etal\ (1996).

\item The typical dust extinction derived using standard extinction
laws is only moderate ($A_V=0.5$--1.0 mags) and very
similar to that inferred in low redshift field galaxies.
If there is a large population of obscured star-forming systems
at $z=1$ they are not common in known samples --- their rate of occurrence
in the CFRS at $z=1$ must be $\ls 10\%$ or they 
would be detected in our data. This limit is 
similar to the results of Pettini \etal\ for the $z>3$ Steidel \etal\
galaxies and consistent with deep sub-mm observations 
which estimate that massive obscured star-forming systems
make up approximately 10\% of high-redshift galaxies (Lilly \etal\ 1998B).

\item A cautionary note is the nature of the dust extinction law: if we
follow the Calzetti attenuation prescription we imply much higher obscuration
of the \Halpha\ line and a star-formation rate at $z=1$ three times higher still. 
It is unclear however whether such a large correction should be applied
to the integrated light of {\em all\/} galaxies at high-redshift.
This issue can only be resolved by further direct measurement of the $\Hbeta/\Halpha$
decrement in these galaxies and the derivation of the extinction to the nebular
regions independent of the stellar UV flux.

\item The mean star-formation rate of a $z=1$ CFRS galaxy is $\sim 20$--60 \Msun\
yr$^{-1}$ (for Salpeter IMF and the range of dust laws we have studied), 
which is enough to make such a $L^*$
galaxy in a few Gyr. This is a factor of several higher than ordinary
spiral galaxies of comparable luminosity today.

\item The large scatter in the distribution of \Halpha\ to UV light
is much better fit by a model in which star-formation
occurs in episodic bursts with intervals of $\sim 0.2$--0.3 Gyr
and of length $\sim 0.1$--0.2 Gyr.
Pure continuous star-formation is strongly ruled out, even with
variable extinction, as a sole explanation.

\item We find qualitative trends for star-forming systems to have
blue colours and peculiar morphology (especially interactions).

\item The star-formation history of the Universe, as inferred from
the Balmer lines, is  qualitatively similar to that inferred from
the UV but corrected upwards by factors of at least 2--3 at all redshifts.
Although the overall form of a rise to $z=1$ is preserved, a compilation
of dust-insensitive data does not yet demonstrate with certainty whether
there is a turnover beyond $z=1$.

\end{enumerate}

Finally, it is clear that the era of detailed spectral studies of
high-redshift galaxies is upon us, made possible by the advent of
intermediate to high-resolution spectrographs on 4m telescopes.  
In the next few years with larger samples and better near-IR
spectrographs on 8\,m class telescopes (many with the ability to
spatially resolve spectra in 2D) it will be possible to model the
spectral evolution of the high-redshift galaxy population in much
greater details. In particular the field will be able to focus much
more on detailed astrophysics (stellar/dust/gas compositions, star-formation
rates, dynamics, etc.) rather than simple statistics.

\section*{ACKNOWLEDGEMENTS}

We are grateful to the UKIRT time assignment committee for their
support of this programme and to the staff and telescope operators of
the Joint Astronomy center for their support. We especially wish to
thank Gillian Wright for technical and astronomical advice. The data
reduction and analysis was performed primarily with computer hardware
supplied by the Anglo-Australian Observatory and STARLINK. Special
thanks go to Laurence Tresse and Francois Hammer of the CFRS team for
making available there raw UV and [OII] data in machine-readable
form. We would also like to thank Jarle Brinchmann and Richard Ellis
for providing the HST images to allow us to explore the role of
morphology.  The interpretation of the data presented here was greatly
aided by helpful discussions with many colleagues, especially 
Bob Abraham, Joss Bland-Hawthorn, Daniela Calzetti, Carlos Frenk, 
Claus Leitherer and Max Pettini. The authors would
also like to thank an anonymous questioner at the 1997 IAU meeting in Kyoto who
prompted the authors into looking in to the detailed modeling of
time-dependent \Halpha\ and UV fluxes.

\vfill\eject

\onecolumn

\begin{table}

\caption{\label{tab-observations}}
{\bf OBSERVED SAMPLE AND FLUX MEASUREMENTS}\\
\\
\begin{tabular}{ccccclccl}
\hline
CFRS\# & Exposure & $z$ & $I_{AB}$ & $(V-I)_{AB}$ & EW [OII] & \multicolumn{1}{c}{$F(\Halpha)$} & $J_C$ & HST Morphology  \\
& (secs) & (mags) & (mags) &  &  (rest \AA) & $10^{-17}$ergs cm$^{-2}$ s$^{-1}$) & (mags) \\
\hline
00.1579 & 1000 &  0.811 & 22.40 &  1.45 & $33 \pm  6$ & $  0.0 \pm   13.9 $ & $ 19.67 \pm  0.13$  & --- \cr
03.0125 & 1000 &  0.790 & 22.08 &  2.08 & $19 \pm  7$ & $ 20.1 \pm    7.8 $ & $ 20.19 \pm  0.10$  & --- \cr
03.0133 & 1500 &  1.048 & 22.45 &  1.00 & $65 \pm 16$ & $ 52.9 \pm   33.9 $ &  $>21.87$           & --- \cr
03.0316 & 1500 &  0.815 & 21.98 &  2.91 & $12 \pm  2$ & $  0.0 \pm    6.8 $ &  $>22.87$           & Spiral\cr
03.0615 &  500 &  1.048 & 22.01 &  0.96 & $27 \pm  5$ & $  0.0 \pm   24.9 $ &  $>20.87$           & --- \cr
03.1534 & 2000 &  0.798 & 22.45 &  0.70 & $39 \pm  8$ & $ 68.9 \pm   23.8 $ & $ 20.70 \pm  0.23$  & --- \cr
10.1220 & 1200 &  0.909 & 22.36 &  0.97 & $20 \pm  4$ & $ 75.5 \pm   16.2 $ &  $>21.69$           & Peculiar (merger)\cr
14.0600 & 4000 &  1.038 & 21.53 &  0.69 & $27 \pm  9$ & $108.2 \pm   19.0 $ & $ 19.08 \pm  0.19$  & Peculiar (close pair)\cr
14.0818 & 1000 &  0.899 & 21.02 &  1.12 & $19 \pm  2$ & $  0.0 \pm   30.5 $ & $ 18.74 \pm  0.11$  & --- \cr
14.1496 & 2400 &  0.899 & 21.80 &  1.13 & $28 \pm  6$ & $  0.0 \pm   15.7 $ & $ 19.10 \pm  0.10$  & Compact \cr
22.0770 & 3000 &  0.819 & 21.78 &  1.58 & $39 \pm  5$ & $ 42.9 \pm    6.4 $ & $ 20.74 \pm  0.18$  & --- \cr
22.1313 & 6000 &  0.819 & 21.74 &  0.84 & $40 \pm  5$ & $ 40.5 \pm    3.6 $ & $ 20.36 \pm  0.08$  & Peculiar (merger)\cr
22.1406 & 4000 &  0.818 & 22.16 &  1.16 & $55 \pm  4$ & $ 78.3 \pm    6.2 $ & $ 21.24 \pm  0.20$  & Compact \cr
\hline
\end{tabular}
\end{table}

\begin{table}
\caption{\label{tab-luminosities}}
{\bf DERIVED LUMINOSITIES AND VELOCITY WIDTHS}\\
\\
\begin{tabular}{ccccc}
\hline
CFRS\# & $L(\Halpha)\,^{\rm (i)}$ & $L([OII])\,^{\rm (i)}$& $L(2800\rm\AA)\,^{\rm (i)}$  
       & FWHM\,$^{\rm (ii)}$\\
       & ($10^{41}$ergs s$^{-1}$) & ($10^{41}$ergs s$^{-1}$) & ($10^{27}$ergs s$^{-1}$ Hz$^{-1}$)
       & rest km$/$sec\\
\hline
00.1579 & $  0.0 \pm    5.2$ & $  3.8 \pm    0.5$ & $ 18.5$ & --- \cr 
03.0125 & $  7.1 \pm    2.8$ & $  6.5 \pm    1.5$ & $ 10.6$ & 457 \cr 
03.0133 & $ 34.7 \pm   22.2$ & $ 12.9 \pm    1.8$ & $ 52.2$ & 392 \cr 
03.0316 & $  0.0 \pm    2.5$ & $  2.4 \pm    0.4$ & $  3.6$ & --- \cr 
03.0615 & $  0.0 \pm   16.3$ & $ 22.6 \pm    2.4$ & $ 81.3$ & --- \cr 
03.1534 & $ 24.8 \pm    8.6$ & $ 13.2 \pm    1.8$ & $ 42.7$ & 283 \cr 
10.1220 & $ 36.2 \pm    7.8$ & $  7.7 \pm    1.4$ & $ 43.5$ & 260 \cr 
14.0600 & $ 69.5 \pm   12.2$ & $ 36.4 \pm    8.3$ & $154.9$ & 377 \cr 
14.0818 & $  0.0 \pm   14.3$ & $ 16.1 \pm    1.9$ & $125.3$ & --- \cr 
14.1496 & $  0.0 \pm    7.4$ & $ 14.9 \pm    1.9$ & $ 60.5$ & --- \cr 
22.0770 & $ 16.3 \pm    2.4$ & $ 13.5 \pm    1.1$ & $ 29.8$ & unresolved \cr 
22.1313 & $ 15.4 \pm    1.4$ & $ 19.5 \pm    2.5$ & $ 70.8$ & 251 \cr 
22.1406 & $ 29.8 \pm    2.4$ & $ 21.8 \pm    1.2$ & $ 33.3$ & 279 \cr 
\hline
\end{tabular}
\par
Notes: 
\begin{enumerate}
\item To correct for dust using the final
extinction values derived in
Section~\ref{sec-sfrz1}
multiply the above values by the following factors:\\
$L(\Halpha)$: $\times 1.6$, $L([OII])$: $\times 2.4$, 
$L(2800\rm\AA)$: $\times 3.1$.
\item The instrumental resolution ranges from 70--100 km$/$s (FWHM) and has
been subtracted in quadrature from these values.
\end{enumerate}
\end{table}

\begin{table}
\caption{\label{tab-conversions}}
{\bf CONVERSIONS OF \Halpha,UV TO STAR-FORMATION RATES}\\
\\
Luminosities are for 1 \Msun\ yr$^{-1}$. 
\\
\\
\begin{tabular}{lclcrrrrrrrrrrr}
\hline
Model & $Z/\Zsun$ & IMF & $L(\Halpha)$ & 
   \multicolumn{3}{c}{$L(1500\rm\AA)$} 
   && \multicolumn{3}{c}{$L(2800\rm\AA)$}  && \multicolumn{3}{c}{$L(2800\hbox{\AA})/L(\Halpha)$}\cr
      &  & & ($10^{41}$ergs s$^{-1}$) 
      & \multicolumn{3}{c}{($10^{27}$ergs s$^{-1}$ Hz$^{-1}$)}
      && \multicolumn{3}{c}{($10^{27}$ergs s$^{-1}$ Hz$^{-1}$)} 
      && \multicolumn{3}{c}{($10^{-14}$ Hz$^{-1}$)}   \cr
      &&&& 0.1 Gyr & 1.0 Gyr & 3.0 Gyr & \phantom{M} & 0.1 Gyr & 1.0 Gyr & 3.0 Gyr 
      && \multicolumn{3}{c}{(1 Gyr)} \cr
\hline
  BC96 (kl96) &   0.02 & SC &  0.68  &  4.36 &{\bf   6.64 }&  7.19  & &  2.84 &  {\bf 6.23} &  8.63 &&&  9.2 \cr
  BC96 (kl96) &   0.20 & SC &  0.58  &  3.93 &{\bf   5.36 }&  5.40  & &  2.84 &  {\bf 5.35} &  6.58 &&&  9.2 \cr
  BC96 (kl96) &   0.40 & SC &  0.50  &  3.80 &{\bf   4.82 }&  4.83  & &  2.89 &  {\bf 4.96} &  5.80 &&&  9.9 \cr
  BC96 (kl96) &   1.00 & SC &  0.40  &  3.48 &{\bf   4.02 }&  4.02  & &  2.93 &  {\bf 4.36} &  4.81 &&&  10.9 \cr
  BC96 (kl96) &   2.50 & SC &  0.28  &  3.07 &{\bf   3.33 }&  3.33  & &  2.93 &  {\bf 3.88} &  4.10 &&&  13.9 \cr
  BC96 (gs95) &   1.00 & SC &  0.61  &  3.64 &{\bf   4.39 }&  4.40  & &  2.39 &  {\bf 3.72} &  4.13 &&&  6.1 \cr
  BC96 (gsHR) &   1.00 & SC &  0.61  &  3.64 &{\bf   4.39 }&  4.40  & &  2.39 &  {\bf 3.72} &  4.13 &&&  6.1 \cr
  PEG (Pad)   &   1.00 & SC &  0.41  &  4.13 &{\bf   4.72 }&  4.72  & &  3.15 &  {\bf 4.80} &  5.39 &&&  11.7 \cr
  PEG (Gen)   &   1.00 & SC &  0.45  &  3.84 &{\bf   4.42 }&  4.43  & &  2.94 &  {\bf 4.57} &  5.18 &&&  10.2 \cr

\\					      
  M98         &   1.00 & SC &        &       &{\bf    3.50}&	           &&&  {\bf  5.10}   \cr
\\					      					
  BC96 (kl96) &   0.02 & SP &  2.23  & 10.21 &{\bf  12.56 }& 12.87  & &  6.29 & {\bf  9.48} & 10.78 &&&  4.2 \cr
  BC96 (kl96) &   0.20 & SP &  1.96  &  9.27 &{\bf  10.93 }& 10.96  & &  6.32 & {\bf  8.85} &  9.53 &&&  4.5 \cr
  BC96 (kl96) &   0.40 & SP &  1.70  &  9.22 &{\bf  10.49 }& 10.50  & &  6.59 & {\bf  8.76} &  9.24 &&&  5.1 \cr
  BC96 (kl96) &   1.00 & SP &  1.35  &  8.61 &{\bf   9.37 }&  9.37  & &  6.85 & {\bf  8.46} &  8.73 &&&  6.3 \cr
  BC96 (kl96) &   2.50 & SP &  0.90  &  7.62 &{\bf   7.99 }&  7.99  & &  6.87 & {\bf  7.94} &  8.07 &&&  8.8 \cr
  BC96 (gs95) &   1.00 & SP &  1.97  &  8.73 &{\bf   9.77 }&  9.77  & &  5.50 & {\bf  7.00} &  7.25 &&&  3.6 \cr
  BC96 (gsHR) &   1.00 & SP &  1.97  &  8.73 &{\bf   9.77 }&  9.77  & &  5.50 & {\bf  7.00} &  7.25 &&&  3.6 \cr
  PEG (Pad)   &   1.00 & SP &  1.19  &  8.64 &{\bf   9.31 }&  9.31  & &  6.16 & {\bf  7.68} &  7.97 &&&  6.5 \cr
  PEG (Gen)   &   1.00 & SP &  1.28  &  8.02 &{\bf   8.66 }&  8.66  & &  5.68 & {\bf  7.16} &  7.46 &&&  5.6 \cr
\\					      					
  K83         &   1.00 & SP-like& 1.12            		                              \cr
  M96         &   1.00 & SP &        &       &{\bf   11.06}&	            &&& {\bf  7.04}   \cr
  M98         &   1.00 & SP & 1.41   &       &{\bf    8.00}&	            &&& {\bf  7.90}   \cr

\hline
\end{tabular}
\\
Notes:\\
\begin{enumerate}
\item SC and SP are the Scalo (1986) and Salpeter (1955) IMFs
\item K83 is Kennicutt's (1983) conversion for a Salpeter-like IMF
\item M96 is Madau \etal\ (1996) values for 0.1-1 Gyr, derived from Bruzual \& Charlot (1993);
M98 are the values from Madau \etal\ (1998)
\item BC96 are values derived from Bruzual \& Charlot (1996) models, multi-metallicity
`kl96' stellar spectra are based on stellar model atmospheres (Lejeune \etal, 1996, Kurucz, 1995)
and `gs95' are based on observed Gunn \& Stryker (1983) spectra. All use the 
`Padova' stellar evolutionary tracks.
\item PEG are values derived from the PEGASE models (Fioc \etal\ 1997)
for the `Padova' and `Geneva' stellar evolutionary tracks respectively.
\end{enumerate}
\end{table}

\begin{table}
\label{lastpage}
\caption{Best fit model parameters\label{tab-likepars}}

\begin{tabular}{|l|l|l|l|}
\hline
Description of model & Best fit parameters with errors, in the form & log($L_{best}$) & Monte-Carlo deviation\\
& $P_{\rm fitted} = 2\sigma < 1\sigma <$ {\bf best value } $< 1\sigma < 2\sigma$ & &  of $L_{best}$ (percentiles)\\
\hline
\hline
Continuous star-formation & $C_{\sigma} = 0.61 < 0.65 < {\bf 0.70  }< - < -$ & $-38.42$ & 98.6 \\
rate \\
\hline
Continuous plus constant & $F = 0.87 < 0.90 < {\bf 1.00  }< - < -$ & $-36.09$ & 87.4 \\
starbursts of fixed mass & $C_{\sigma}$ not relevant because $F_{best} = 1.0$ & \\
& $M_{\mu} = 1.23 < 1.31 < {\bf 1.39  }< 1.48 < 1.57 \times 10^{9}$ \Msun\ & \\
& $\tau = - < 50.6 < {\bf 80.0 } < 88.4 < 97.8$ Myr & \\
\hline
Continuous plus constant & $F = 0.93 < 0.96 < {\bf 1.00  }< - < -$ & $-35.88$ & 82.0 \\
starbursts of variable mass & $C_{\sigma}$ not relevant because $F_{best} = 1.0$ & \\
& $M_{\mu} = 1.57 < 1.84 < {\bf 2.15  }< 2.56 < 3.03 \times 10^{9}$ \Msun\ & \\
& $M_{\sigma} = 0.18 < 0.22 < {\bf 0.30  }< 0.32 < 0.34$ & \\
& $\tau = 65.9 < 72.6 < {\bf 80.0 } < 107.0 < 143.1$ Myr & \\
\hline
Continuous plus exponential & $F = 0.83 < 0.94 < {\bf 1.00  }< - < -$ & $-36.21$ & 88.4 \\
starbursts of fixed mass & $C_{\sigma}$ not relevant because $F_{best} = 1.0$ & \\
& $M_{\mu} = 4.28 < 4.71 < {\bf 5.18  }< 5.54 < 5.93 \times 10^{9}$ \Msun\ & \\
& $\tau = 122.7 < 140.1 < {\bf 160.0  }< 170.5 < 181.7$ Myr & \\
\hline
Continuous plus exponential & $F = 0.92 < 0.96 < {\bf 1.00 } < - < -$ & $-36.04$ & 88.0 \\
starbursts of variable mass & $C_{\sigma}$ not relevant because $F_{best} = 1.0$ & \\
& $M_{\mu} = 2.67 < 2.91 < {\bf 3.16  }< 3.37 < 3.60 \times 10^{9}$ \Msun\ & \\
& $M_{\sigma} = - < - < {\bf 0.10 } < 0.13 < 0.17$ & \\
& $\tau = 65.0 < 72.1 <{\bf  80.0 } < 106.3 < 141.4$ Myr & \\
\hline
\end{tabular}
\bigskip\\
Notes: 
\begin{enumerate}
\item Parameter ranges `$1\sigma$' and `$2\sigma$' correspond to $\Delta\log{\cal L} = -0.5$, $-1.0$.
\item $M_{\sigma}$ is expressed as a fraction of  $M_{\mu}$.
\end{enumerate}

\end{table}

\end{document}